\documentclass[reprint,10pt,prl,aps,superscriptaddress,twocolumn,longbibliography]{revtex4-1}
\usepackage{amssymb,amsmath}
\usepackage{graphicx}
\usepackage{color}
\usepackage{natbib}
\usepackage{hyperref}
\usepackage[utf8]{inputenc}
\usepackage{epstopdf}

\begin{document}
\makeatletter
\newenvironment{sqcases}{%
  \matrix@check\sqcases\env@sqcases
}{%
  \endarray\right.%
}
\def\env@sqcases{%
  \let\@ifnextchar\new@ifnextchar
  \left\lbrack
  \def\arraystretch{1.2}%
  \array{@{}l@{\quad}l@{}}%
}
\makeatother

\newcommand{\e}{{\rm e}}
\newcommand{\rmi}{{\rm i}}
\renewcommand{\Im}{\mathop\mathrm{Im}\nolimits}
\newcommand{\red}[1]{{\color{red}#1}}
\newcommand{\blue}[1]{{\color{blue}#1}}
\newcommand{\green}[1]{\textcolor[rgb]{0,0.7,0.4}{#1}}
\newcommand{\commentSasha}[1]{{\color{blue}{\bf Sasha:}\it #1}}

\renewcommand{\cite}[1]{[\onlinecite{#1}]}
\newcommand{\cra}[1]{\hat{a}^{\dag}_{#1}}  
\newcommand{\ana}[1]{\hat{a}_{#1}^{\vphantom{\dag}}}         
\newcommand{\num}[1]{\hat{n}_{#1}}         
\newcommand{\ket}[1]{\left|#1\right>}      
\newcommand{\bra}[1]{\left<#1\right|}
\newcommand{\eps}{\varepsilon}      
\newcommand{\om}{\omega}      
\newcommand{\kap}{\varkappa}      

\newcommand{\spm}[1]{#1^{(\pm)}}
\newcommand{\skvk}[2]{\left<#1\left|\frac{\partial #2}{\partial k}\right.\right>} 
\newcommand{\skvv}[2]{\left<#1\left|#2\right.\right>}
\newcommand{\df}[1]{\frac{\partial #1}{\partial k}}
\newcommand{\ds}[1]{\partial #1/\partial k}


\title{Topological interface states mediated by  spontaneous symmetry breaking
}

\author{Roman~S.~Savelev}
\affiliation{ITMO University, Saint Petersburg 197101, Russia}

\author{Maxim~A.~Gorlach}
\affiliation{ITMO University, Saint Petersburg 197101, Russia}

\author{Alexander N. Poddubny}
\affiliation{ITMO University, Saint Petersburg 197101, Russia}
\affiliation{Ioffe Institute, Saint Petersburg 194021, Russia}
\email{poddubny@coherent.ioffe.ru}

\begin{abstract}

We propose  a one-dimensional nonlinear system of coupled anharmonic oscillators that  dynamically undergoes a topological transition switching from the {disordered} and topologically trivial  phase into the nontrivial one due to the spontaneous symmetry breaking. The topological transition is accompanied by the formation of the topological interface state in the spectrum of linearized excitations of the stationary phase. Our findings thus highlight the potential of the nonlinear systems for hosting the topological phases and uncover a fundamental link between the spontaneous symmetry breaking mechanism and topological edge states.

\end{abstract}

\maketitle

{Spontaneous symmetry breaking in nonlinear systems is one of pivotal concepts of modern physics which has important implications for high-energy physics~\cite{Higgs,Chatrchyan}, physics of condensed matter~\cite{Yang}, nonlinear optical systems~\cite{Herring}, Bose-Einstein condensates~\cite{Ostrovskaya,Aleiner} and metamaterials~\cite{MingkaiLiu}}. Spontaneous symmetry breaking in condensed matter system is often accompanied by a second-order phase transition, e.g. from a paramagnetic to a ferromagnetic state~\cite{landau5}.

An interesting question is whether it is possible to realize a topological phase transition from a trivial to a gapped nontrivial phase. One could expect that the resulting spontaneously broken phase would host {topologically nontrivial} linear excitations \cite{bernevig2013}. Topological edge or interface states of electrons~\cite{bernevig2013}, light~\cite{Lu2014,Lu2016,Khanikaev-NP} and sound~\cite{Huber2016} have recently received much attention due to their prospects for realization of disorder-robust one-way  transport of information. Presently, the interest is shifting towards topological states in nonlinear and interacting systems promising higher tunability and richer fundamental physics~\cite{Hadad,Hadad-ACS,Liew2016,Solnyshkov2016,Solnyshkov,Leykam2016,Chong-NJP,Liberto,Gorlach-2017}.
However, there is still no clear recipe to realize a nonlinear system with  edge or interface states between
topologically distinct domains appearing due to spontaneous symmetry breaking.  
Harnessing spontaneous nature of the transition would ensure dynamical and low-energy-cost  reshaping of topologically trivial {potential landscapes into the nontrivial ones and vice versa}.

{In this Letter}, we examine a spontaneous formation of interface excitations in the linearized spectrum of the periodic {array} of nonlinear mechanical oscillators with anharmonic repulsive coupling. We show, that after the Peierls-like symmetry-breaking transition~\cite{braun2004frenkel} an initially disordered system can form metastable topologically distinct regions with linear topological edge states localized at the domain walls. In another words, the repulsion-induced symmetry breaking gaps the spectrum and generates  the topological interface states from the disorder.

The proposed mechanism is qualitatively different from the formation of topological solitons~\cite{manton2007topological}, i.e. boundaries between topologically distinct phases,
studied across different domains ranging from the early Universe physics \cite{Kibble1976} to liquid helium~\cite{Zurek1985}, liquid crystals \cite{Yurke1991} and Bose-Einstein condensates~\cite{Lamporesi2013,Solnyshkov2016,Solnyshkov}. Namely, we aim for  topological 
edge states in the band gap centered at non-zero frequency. These are in stark contrast both with the stationary topological solitons and with the zero-frequency localized modes of the linearized spectrum of topological solitons~\cite{manton2007topological}.

Considered  spontaneous nonlinear interface state formation in an initially symmetric system is also distinct from one occurring in an intrinsically asymmetric nonlinear system~\cite{Hadad,Hadad-ACS}. It is also different from the case where the potential topology is imprinted  by an external inhomogeneous pump~\cite{Peano2015,Liew2016} or magnetic field~\cite{Nalitov2015}. 

The proposed  concept  is quite general and has implications beyond nonlinear mechanical oscillator arrays~\cite{Sato2006,Prodan2017} providing insights into the physics of zigzag and helical cold ions arrays~\cite{Landa2013,Nigmatullin2016}, buckled mechanical~\cite{Stoop2015,Paulose2015,Bertoldi2017} and optomechanical~\cite{Taylor2017} structures, bifurcations in superconducting circuits~\cite{Engelhardt}, nanowires~\cite{Cheon2015,Komsa2017}  and nonlinear quantum optics~\cite{Liberto,Gorlach-2017}.

\begin{figure}[t]
	\center{\includegraphics[width=1\columnwidth]{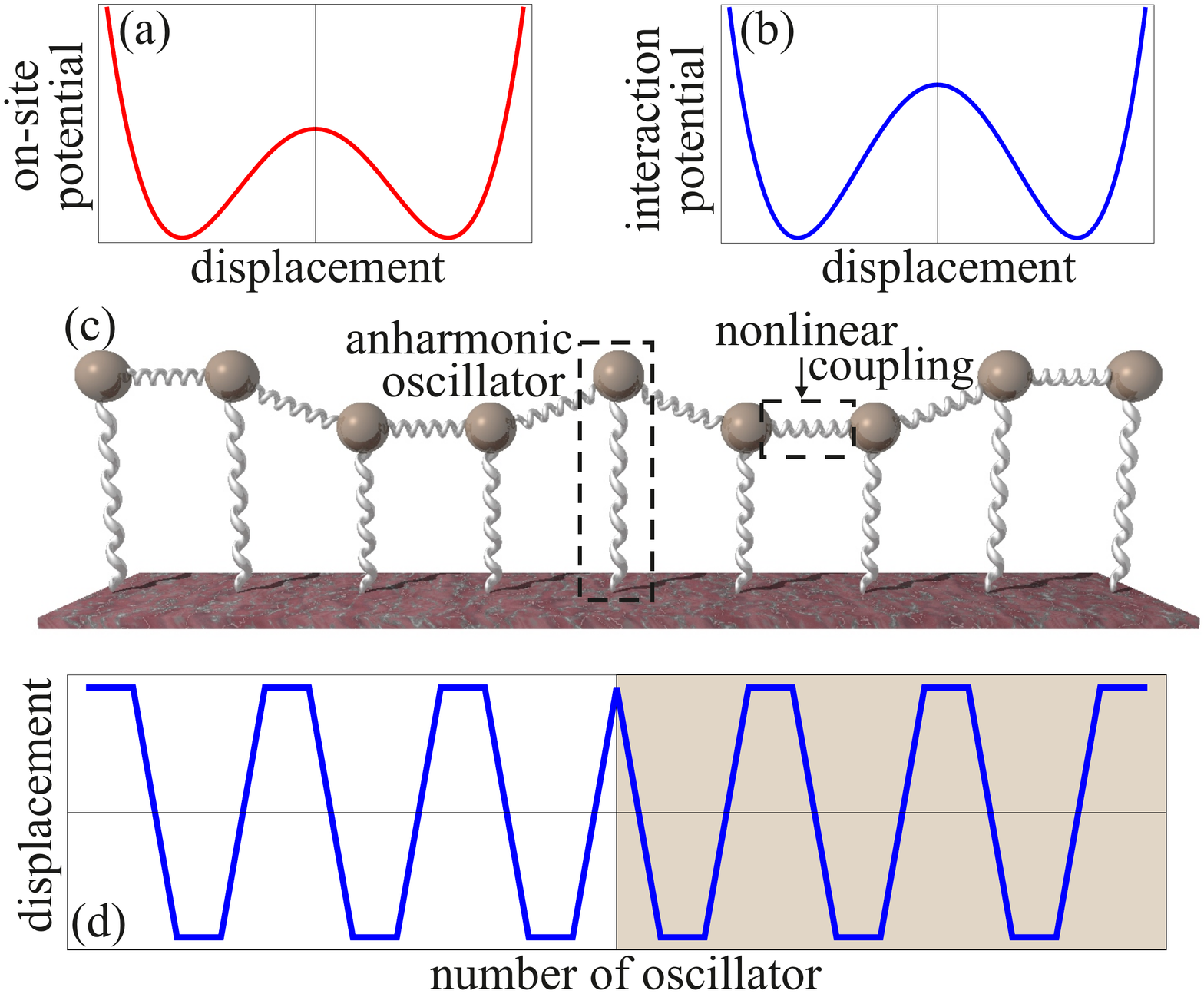}}
	\caption{An array of anharmonic oscillators with nonlinear coupling between them. (a) On-site potential for the individual oscillator. (b) Interaction potential for two neighboring oscillators. (c) A schematic of the system under study. (d) Stationary states of the oscillators after sufficiently long evolution time  (tetramerized stationary state) with a domain wall present.}
	\label{fig:Scheme}
\end{figure}

 The considered system (Fig.~\ref{fig:Scheme}) is based on the array of identical anharmonic oscillators [Fig.~\ref{fig:Scheme}(c)] with double-well on-site potential~[Fig.~\ref{fig:Scheme}(a)] and anharmonic coupling between the nearest neighbors~[Fig.~\ref{fig:Scheme}(b)]. The entire array is described by potential function
\begin{equation}\label{eq:U}
U = \sum\limits_{n=1}^{N} (a_2y_n^2 + a_4y_n^4) + \sum\limits_{n=1}^{N-1} [b_2(y_n - y_{n+1})^2 + b_4(y_n - y_{n+1})^4]\:,
\end{equation}
where $a_2$, $a_4$ and $b_2$, $b_4$ are on-site and inter-site force constants, respectively. The terms $\propto a_4$ and $\propto b_4$ describe the anharmonicity of the potential.

Our analysis reveals that one of the stable stationary states of such system is the {\it tetramer stationary state} with the period of 4 when stationary displacements of oscillators satisfy the conditions 
\begin{equation}\label{Conditions}
y_{4n+1}^{(0)}=y_{4n+2}^{(0)}=-y_{4n+3}^{(0)}=-y_{4n+4}^{(0)}=v_{0}
\end{equation}
as schematically sketched in Fig.~\ref{fig:Scheme}(d). The linearized spectrum of small oscillations in the vicinity of this stationary state reproduces the Su-Schrieffer-Heeger model (SSH) describing tunneling-coupled  arrays with alternating strong and weak tunneling links~\cite{bernevig2013,Shen,Malkova:09,Schomerus:13,Slob,Zhang2015,Engelhardt,Nalitov2017}. Hence, the linear spectrum contains interface states localized at the domain wall.

Besides tetramer stationary states, the system supports also {\it monomer} and {\it dimer} stationary states with $y_n=\text{const}$ and $y_{2n+1}=-y_{2n}=u$, respectively. However, as detailed in Supplemental Materials, Secs.~S1, S2, small oscillations in the vicinity of these stationary states do not reproduce the physics of SSH model and do not yield any topological states.

Consequently, the tetramer structure is the simplest mechanical realization of the spontaneously induced topological interface states. The stationary displacements $v_0$ in tetramer state are found from the condition $\partial\,U/\partial\,y_n=0$:
\begin{equation}\label{StatDisplacement}
v_0=\pm \sqrt{-\dfrac{a_2+2\,b_2}{2\,a_4+16\,b_4}}\:.
\end{equation}
The stability condition imposes an additional inequality on the second derivatives of the potential function (see Sup. Mat., Sec. S3). Further, to ensure that the tetramer stationary state still persists even in the case of a finite array, one more additional condition should be fulfilled (Sup. Mat., Sec. S2): 
\begin{equation}\label{eq:rel_coefs}
\dfrac{b_2}{a_2}=4\dfrac{b_4}{a_4}\:.
\end{equation}
Note that with the latter condition the stationary displacement is the same as for the single anharmonic oscillator: $v_0=\pm\sqrt{-a_2/(2\,a_4)}$.

\begin{figure}[b]
	\center{\includegraphics[width=1\columnwidth]{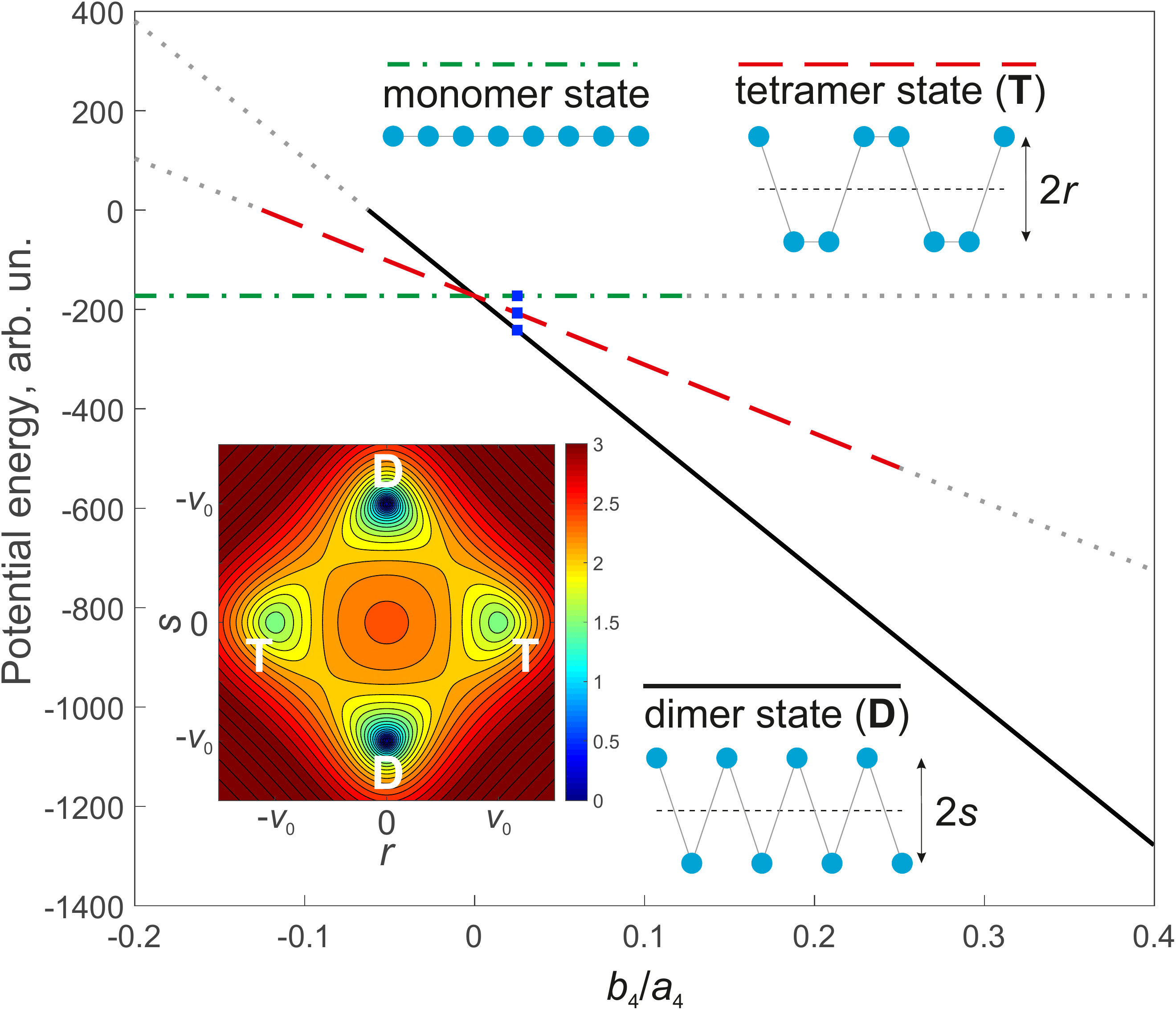}}
	\caption{Potential energy of the stationary states for the array of anharmonic oscillators as a function of the ratio $b_4/a_4$ with fixed parameters $a_2=-12$, $a_4=5/6$; $b_2$ is determined from the condition~(\ref{eq:rel_coefs}). Solid black, dashed red and dash-dotted green lines correspond to dimer, tetramer and monomer stationary states, respectively. Dotted grey lines indicate the values of $b_4/a_4$, where the corresponding stationary states are unstable. Blue squares mark the parameters used in the following calculations. {Inset: potential energy as a function of relative displacements $r$ and $s$  defined in text.}}
	\label{fig:U}
\end{figure}

Tuning the anharmonicity of on-site and coupling potentials given by the coefficients $a_4$ and $b_4$ enables one to change the ratio between the energies of monomer, dimer and tetramer state as indicated in Fig.~\ref{fig:U}, thus defining the global energy minimum. As a representative example, we choose $a_2=-12$, $b_2=-1.2$, $a_4=5/6$ and define $b_4=1/48$ according to Eq.~\eqref{eq:rel_coefs}. The energies of the monomer, dimer and tetramer states in such case are indicated by blue squares in Fig.~\ref{fig:U}. For these parameters both on-site and coupling nonlinearities are described by double-well potentials as depicted in Fig.~\ref{fig:Scheme}(a,b). To further visualize the complicated potential landscape and the interplay between the stationary states, we plot the potential energy for a special class of states given by the equations $y_{4n+1}=r+s$, $y_{4n+2}=r-s$, $y_{4n+3}=-r+s$, $y_{4n+4}=-r-s$ and characterized by only two parameters $r$ and $s$. The calculated color map of the potential function shown as inset in Fig.~\ref{fig:U} features two pairs of local potential minima: {$r=0$, $s=\pm u$ and $s=0$, $r=\pm v_0$} which correspond to dimer and tetramer stationary states, respectively.

As a next step, we consider small oscillations in the vicinity of the tetramer stationary state. We substitute $y_n=y_n^{(0)}+z_n$ into the equations of motion with $y_n^{(0)}$ being the oscillator stationary displacement given by Eq.~\eqref{Conditions} and $z_n$ representing small deviation from the equilibrium state. Keeping only terms linear in $z_n$, we get
\begin{align*}
(\omega^2 - \omega_0^2+2i\,\omega\,\gamma)\,z_n &= -J_1\,z_{n+1} - J_2\,z_{n-1}\;\;\text{(odd sites)},\\
(\omega^2 - \omega_0^2+2i\,\omega\,\gamma)\,z_n &= -J_2\,z_{n+1} - J_1\,z_{n-1}\;\;\text{(even sites)},\nonumber
\end{align*}
with $J_1 = 2\,b_2$, $J_2 = 2\,b_2 + 48\,b_4\,v_0^2$ so that for the chosen parameters $J_2/J_1=-2$. The ``eigenfrequency'' $\omega_{0}$ is given by a sum of on-site and inter-site contributions:
\begin{equation}\label{Eigenfreq}
\omega_{0}^2 = 2\,a_2+4\,b_2+(12\,a_4+48\,b_4)\,v_0^2=\omega_{\rm{site}}^2+J_1+J_2\:.
\end{equation} 
Thus, in terms of bulk properties, oscillations in the vicinity of the tetramer stationary state are captured by the SSH model with alternating links $J_{1}$ and $J_{2}$. The equation for the edge oscillator is similar:
\begin{equation}\label{EdgeOscillator}
\left(\omega^2-\omega_{0e}^2+2i\,\omega\,\gamma\right)\,z_1=-J_1\,z_2\:,
\end{equation}
but the eigenfrequency $\omega_{0e}$ appears to be modified: $\omega_{0e}^2=\omega_0^2-J_2$ which is a consequence of the fact that the edge oscillator has less neighbors. For that reason, even if the array is terminated at the weak link edge, the detuning of the edge oscillator is so large that the edge state is impossible [Sup.~Mat., Sec.~S2].  This is different from the conventional SSH case~\cite{Shen}.

\begin{figure}[t]
	\center{\includegraphics[width=1\columnwidth]{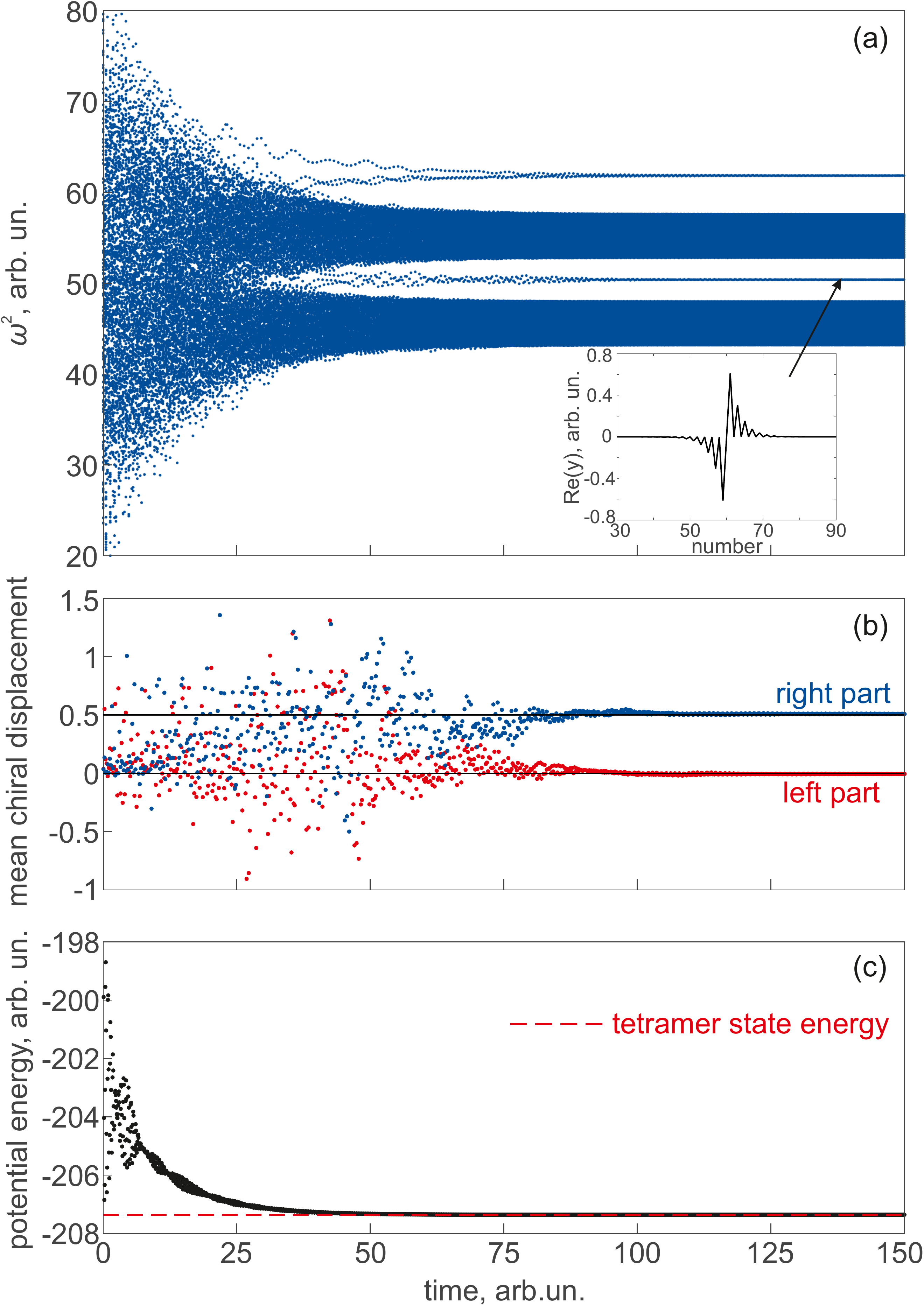}}
	\caption{(a-c) Time evolution of the array of 120 oscillators. (a) Spectrum of small oscillations. Inset shows the  distribution of displacements for the interface mode. (b) Mean chiral displacement calculated for the left (red dots) and right (blue dots) tetramer domains, see Fig.~\ref{fig:Scheme}(d). (c) Potential energy of the system versus evolution time. Red dashed line indicates the potential energy of the tetramer stationary state.}
	\label{fig:Spectrum_MCD_U}
\end{figure}

On the other hand, the topological interface state at the domain wall between the two arrays with the opposite dimerizations is still possible in the geometry of Fig.~\ref{fig:Scheme}(d) [Sup.~Mat., Sec.~S2].
In order to probe the emergence of the topological order and topological interface state, we analyze the dynamics of $N=120$ oscillators by directly solving full dynamic equations with small friction term $\gamma=0.05$ included for convergence. Initial displacements and velocities of the oscillators were randomly distributed in the range $\pm(v_0-\delta y,v_0 + \delta_y)$ and $(-\dot\delta_y, \dot\delta_y)$, respectively, with $v_{0}\approx2.683$ and the maximum deviations $\delta y=0.4$ and {$\dot\delta_y=1.5$}. At each moment of time $t$ the calculated $y_n(t)$ were considered as stationary displacements, and the spectrum of small oscillations was evaluated. In this way we recovered the evolution of the spectrum presented in Fig.~\ref{fig:Spectrum_MCD_U}(a). It is seen that during the evolution the spectrum of the system becomes gapped, and two edge states appear, the midgap  state and the state above the allowed bands. We are interested in the former state, corresponding to the topological zero-energy state in the SSH model~\cite{Shen}. The calculated displacement distribution depicted as inset in Fig.~\ref{fig:Spectrum_MCD_U}(a) confirms that this state is indeed localized at the interface.

Even more exciting feature is the dynamical emergence of the  topological characteristics for  initially disordered system approaching the equilibrium. Quite importantly, the traditional approaches to the topological characterization, for instance Zak phase technique~\cite{Zak} are not applicable here since the system is not strictly periodic at arbitrary moment of time. To circumvent this difficulty, we have adopted the technique of Refs.~\cite{Cardano,Maffei} based on random quantum walks and characterizing the system topology in terms of {\it mean chiral displacement}. 
At each moment of time $t$ we linearize the disordered system and characterize it by an effective tight-binding Hamiltonian, determining the evolution from $t$ to  $t+\tau$.
Next, we calculate the limit of {mean chiral displacement } MCD$(t,\tau)$ of an initially localized state at $\tau\to \infty$, see Sup. Mat., Sec. S4. 

The independently obtained values of MCD$(t,\tau\to \infty)$ for the left and right halves of the array  (before and after the domain wall)  are presented in Fig.~\ref{fig:Spectrum_MCD_U}(b). While at short timescales the results for both halves  are roughly the same and fluctuate with time enormously, after long-time evolution mean chiral displacements converge to 0 and 0.5, which are the values characteristic to the SSH array with different dimerizations~\cite{Cardano}. The obtained values of mean chiral displacement prove the topological origin of the interface state. An interesting additional observation evident from Figs.~\ref{fig:Spectrum_MCD_U}(a,b) is that the topology measured by the mean chiral displacement ``emerges'' not  when the spectrum of the system becomes gapped or potential energy reaches local minimum [$t \gtrsim 30$, Fig.~\ref{fig:Spectrum_MCD_U}(c)], but only  after the in-gap  interface edge states finally stabilizes, $t\gtrsim 100$.

An insightful visualization of the system spectrum both in real and reciprocal space is provided by the density of states (DOS) technique. Real-space- and reciprocal-space-resolved densities of states are calculated as a sum over all eigenstates of the system with a weight that depends on the energy detuning between the energy variable $\omega^2$ and the energy of the $m$th eigenstate $\omega_{0,m}$:
\begin{align*}
&\mathrm{DOS}_{\rm{real}}(\omega^2,n,t) = \\
&\sum\limits_{m=1}^N \dfrac{|y_{m}(n,t)|}{\sqrt{2\pi}\sigma}\exp{\left(-\dfrac{(\omega_{0,m}^2-\omega^2)^2}{2\sigma^2}\right)},\\
&\mathrm{DOS}_{\rm{reciprocal}}(\omega^2,k,t) = \\
&\sum\limits_{m=1}^N \dfrac{|\hat{y}_{m}(k,t)|}{\sqrt{2\pi}\sigma}\exp{\left(-\dfrac{(\omega_{0,m}^2-\omega^2)^2}{2\sigma^2}\right)},
\end{align*}
where $y_{m}$ are  eigenvectors of the linearized system  at a certain moment~$t$, $\hat{y}_{m}$ is the Fourier transform of $y_{m}$, and $\sigma$ is an auxiliary parameter taken as 0.2 in our calculations.\par

The calculated real-space- and reciprocal-space-resolved densities of states at the beginning and in the end of evolution are presented in Fig.~\ref{fig:DOS}. Full time dynamics can be seen in the Supplemental Movies. At $t=0$  the system is disordered, its eigenmodes are delocalized [Fig.~\ref{fig:DOS}(a)] and the spectrum has no band gap [Fig.~\ref{fig:DOS}(b)], in agreement with
Fig.~\ref{fig:Spectrum_MCD_U}(a).
 Examining the density of states in the real space at large evolution times [Fig.~\ref{fig:DOS}(c)], we observe that spectrum of the system becomes gapped, and the interface state localized in the middle of bandgap is formed. Density of states in the reciprocal space [Fig.~\ref{fig:DOS}(d)] provides clues about the dispersion of the bulk bands which closely resembles that in the SSH model further highlighting the topological nature of the studied system.

\begin{figure}[t!]
\center{\includegraphics[width=1\columnwidth]{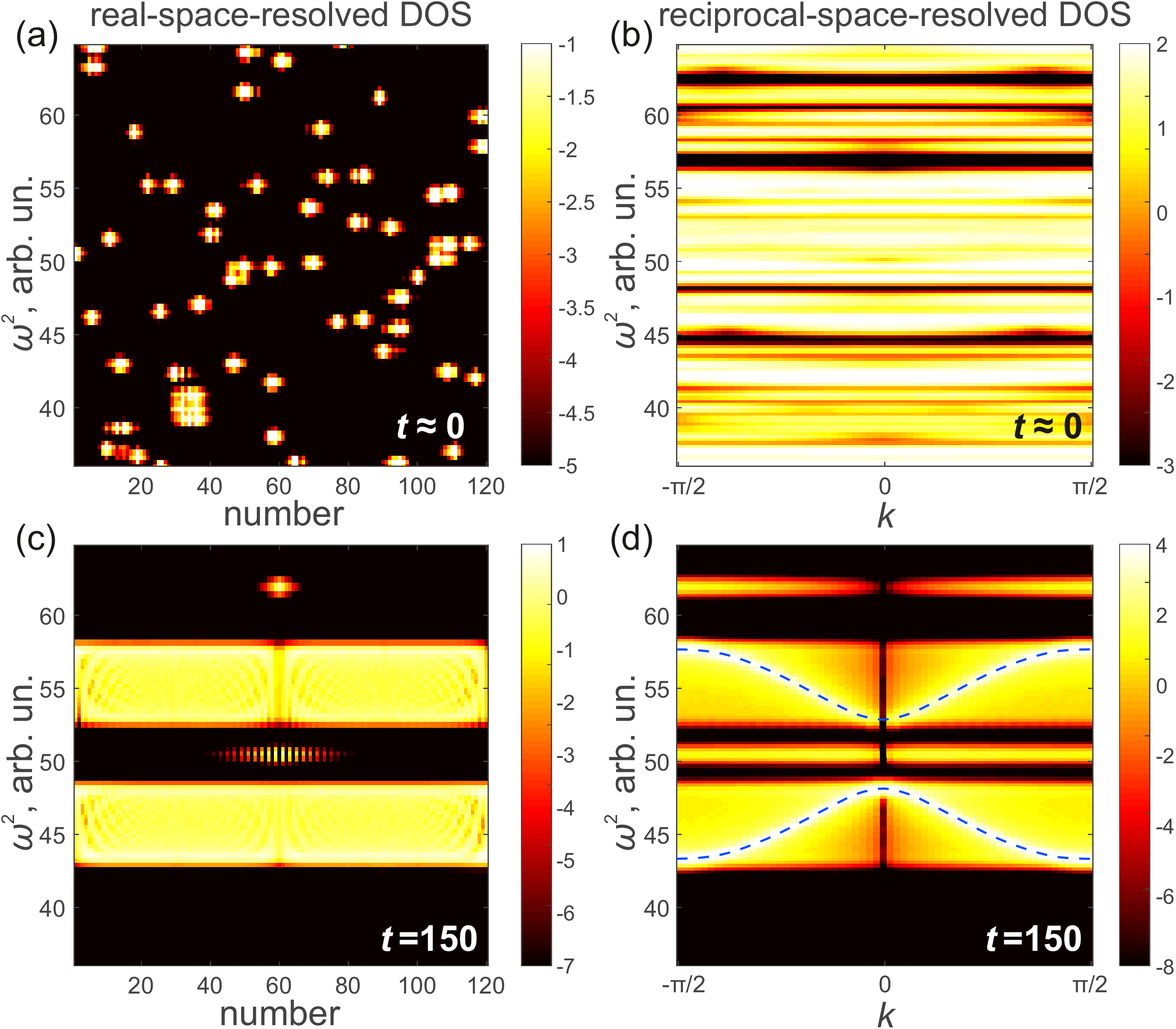}}
\caption{(a) Real space- and (b) reciprocal space-resolved density of states at the beginning of evolution. (c,d) The same as (a,b) for tetramer stationary state at $t=150$~arb.~un. {Dashed curves in (d) mark the maximum density of states.}}
\label{fig:DOS}
\end{figure}

To summarize, our findings prove that nonlinear systems can dynamically switch  from the disordered regime to the  regime with non-zero-frequency topological edge states due to the spontaneous symmetry breaking mechanism. We believe that the fundamental link between spontaneous symmetry breaking and dynamical topological states demonstrated here on a simple example of mechanical system is much more general being applicable to a wide variety of nonlinear electronic, photonic and atomic systems.

{\it Acknowledgments.~--~}Useful discussions with Yu.S.~Kivshar and A.V.~Yulin are gratefully acknowledged. This work was supported by the Russian Science Foundation (Grant No. 16-19-10538). ANP acknowledges partial support by the Foundation for the Advancement
of Theoretical Physics ``Basis''  and the Russian President Grant No.~MD-5791.2018.2.

%


\setcounter{figure}{0}
\renewcommand{\thefigure}{S\arabic{figure}}
\renewcommand{\theequation}{S\arabic{equation}}
\renewcommand{\thesection}{S\arabic{section}}

\onecolumngrid
\appendix
\newpage
\section*{Supplementary Information}
\section{Model, stationary states, linearized equations for small oscillations}

We consider a chain of $N$ nonlinearly coupled anharmonic oscillators with the potential in the form:
\begin{equation}
U = \sum\limits_{i=1}^{N} [a_2y_n^2 + a_4y_n^4] + \sum\limits_{i=1}^{N-1} [b_2(y_n - y_{n+1})^2 + b_4(y_n - y_{n+1})^4].
\end{equation}
The evolution of such system is described with the following equations of motion:
\begin{eqnarray}
\begin{cases}
\medskip
\ddot y_1 + 2\gamma \dot y_1 = -\dfrac{\partial U}{\partial y_1} = -2a_2y_1 - 4a_4y_1^3 - 2b_2(y_1 - y_2) - 4b_4(y_1 - y_2)^3,\\\medskip
\ddot y_n + 2\gamma \dot y_n = -\dfrac{\partial U}{\partial y_n} = -2a_2y_n - 4a_4y_n^3 - 2b_2(2y_n - y_{n-1} - y_{n+1}) - 4b_4[(y_n - y_{n-1})^3 + (y_n - y_{n+1})^3],\\
\ddot y_N + 2\gamma \dot y_N = -\dfrac{\partial U}{\partial y_N} = -2a_2y_N - 4a_4y_N^3 - 2b_2(y_N - y_{N-1}) - 4b_4(y_N - y_{N-1})^3.
\end{cases}
\label{dynamics_eq}
\end{eqnarray}
The stationary states of the system can be found from the system of equations~(\ref{dynamics_eq}) with zero left side. It can be shown that  single nonlinear oscillator possesses two symmetric stationary solutions $v_0 = \pm\sqrt{-\dfrac{a_2}{2a_4}}$, while in an infinite chain of the nonlinear oscillators one can construct different types of solutions, such as:\par
Monomer state: $y_n^{(0)} = v_M = \pm\sqrt{-\dfrac{a_2}{2a_4}}$ with the potential energy per one oscillator $U_M = \dfrac{a_2}{2}v_M^2$.\par
Dimer state: $y_{2n}^{(0)} = v_D = \pm\sqrt{-\dfrac{a_2 + 4b_2}{2(a_4+16b_4)}}$, $y_{2n+1}^{(0)} = -v_D$ with the energy $U_D =  \dfrac{a_2+4b_2}{2}v_D^2$.\par
Tetramer state: $y_{4n}^{(0)} = y_{4n+1}^{(0)} = v_T = \pm\sqrt{-\dfrac{a_2 + 2b_2}{2(a_4+8b_4)}}$, $y_{4n+2}^{(0)} = y_{4n+3}^{(0)} = -v_T$ with the energy $U_T = \dfrac{a_2+2b_2}{2}v_T^2$.\\
The conditions of existence and stability of these solutions and the corresponding constraints on the parameters of the system are discussed in the last section.\par

\textbf{Small oscillations}: Now let us consider small oscillations $z_n(t)$ near a stationary state $y_{n}^{(0)}$ of the system. The system of linear equations for amplitudes $z_n$ can be written down in the frequency domain, after substituting $y_n(t) = y_{n}^{(0)} + z_n\exp(-i\omega t)$ into the system of equations~(\ref{dynamics_eq}) and linearizing it. We obtain one equation for the bulk oscillators:
\begin{equation}
(\omega^2 - \omega_{n_R}^2 - J_{n1} - J_{n2})z_n + J_{n1}z_{n+1} + J_{n2}z_{n-1}  = 0,\\
\label{linear_eq_bulk}
\end{equation}
and two equations for the edge oscillators:
\begin{eqnarray}
\begin{cases}
(\omega^2 - \omega_{1_R}^2 - J_{11})z_1 + J_{11}z_2 = 0,\\
(\omega^2 - \omega_{N_R}^2 - J_{N2})z_N + J_{N2}z_{N-1}  = 0,\\
\end{cases}
\label{linear_eq_edge}
\end{eqnarray}
where $\omega_{n_R}^2 = 2(a_2 + 6a_4(y_{n}^{(0)})^2)$, $J_{n1} = 2(b_2+6b_4(y_{n+1}^{(0)} - y_{n}^{(0)})^2)$, $J_{n2} = 2(b_2+6b_4(y_{n-1}^{(0)} - y_{n}^{(0)})^2)$. Analyzing the linearized system written for an infinite chain [i.e. equation~(\ref{linear_eq_bulk})] one can conclude that the monomer and dimer solutions are characterized by equal squares of the displacements difference $(y_{n+1}^{(0)} - y_{n}^{(0)})^2 = (y_{n-1}^{(0)} - y_{n}^{(0)})^2$ and consequently equal interaction constants $J_{n1} = J_{n2}$. Therefore, the linearized equations of motion describe a system of equally coupled identical linear oscillators with a single dispersion band. On the other hand the interaction constants in the case of tetramer state differ by $|J_{n1} -J_{n2}| = 48b_4v_T^2$, which makes this model similar to the Su-Shrieffer-Heeger one.

\section{Edge and interface states}
\begin{figure}[h]
	\center{\includegraphics[width=0.9\columnwidth]{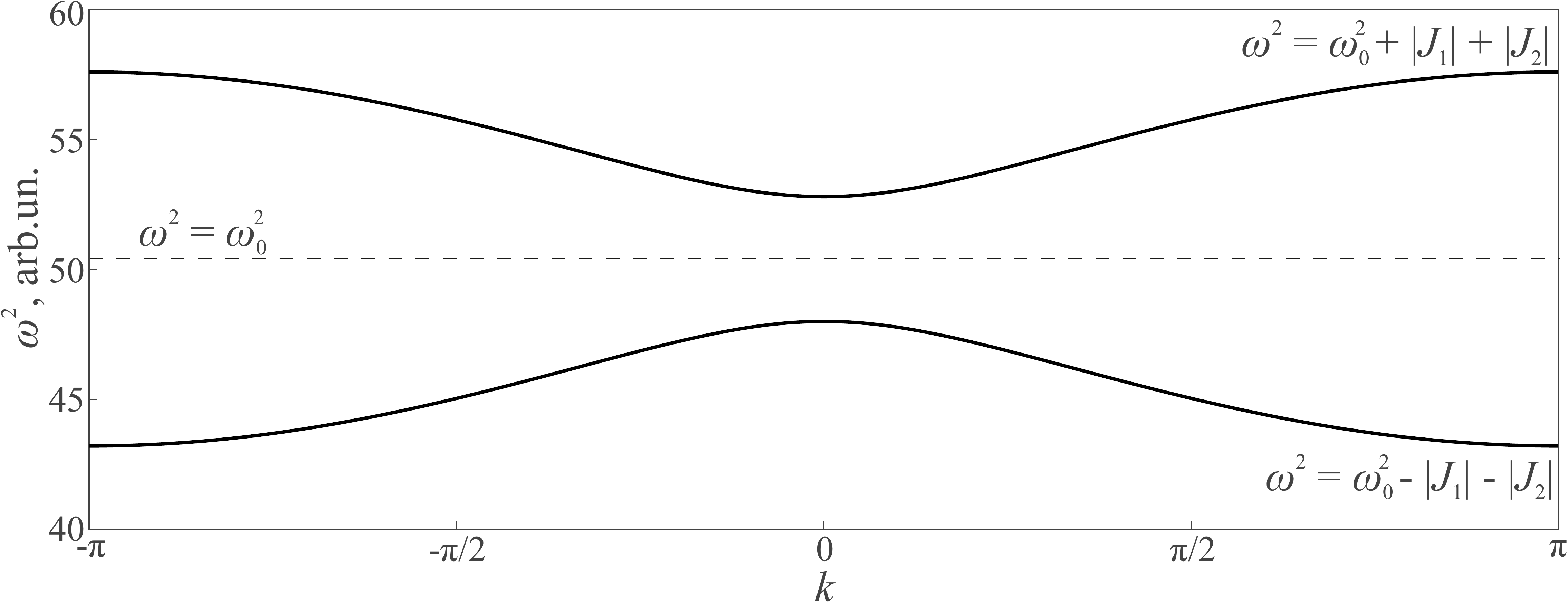}}
	\caption{Linear spectrum --- $\omega^2$ as a function of normalized wavenumber $k$ --- of an infinite chain in the tetramer state. Dashed line marks the center of the bandgap. Following parameters were used in calculations: $a_2=-12$, $a_4=5/6$, $b_2=-6/5$, $b_4=1/48$.}
	\label{Disp_tetramer}
\end{figure}

\begin{figure}[t]
	\center{\includegraphics[width=0.7\columnwidth]{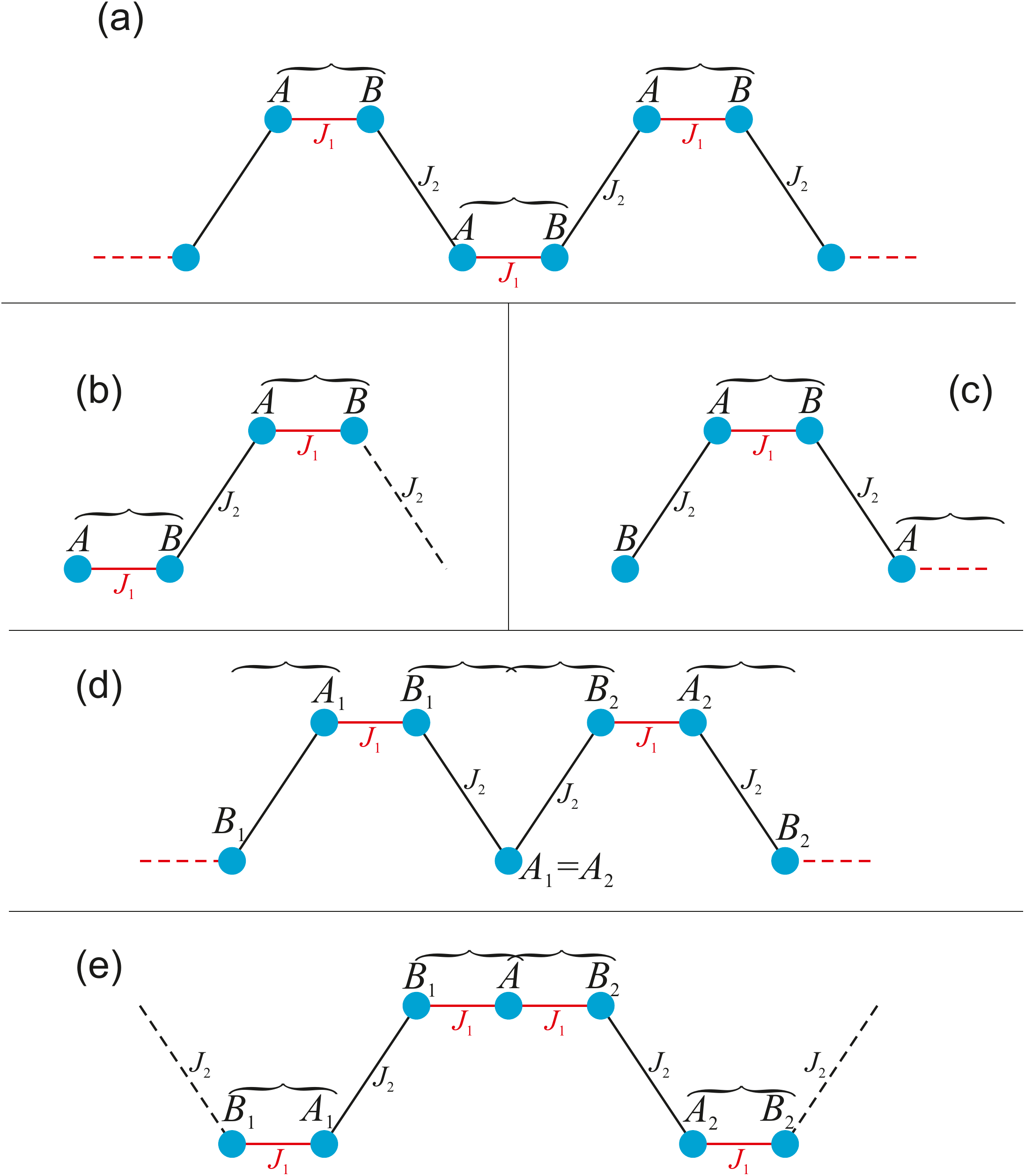}}
	\caption{(a) Scheme of an infinite chain in the tetramer state. (b,c) Scheme of finite chains in tetramer state with two possible edge terminations. (d,e) Scheme of two possible interface configurations between two semi-infinite chains in different tetramer phases. The brackets indicate the unit cell choice in left and right domains. }
	\label{Suppl_scheme}
\end{figure}

Next we consider the behaviour of small ocillations near the tetramer stationary state in more detail. To obtain a linear dispersion of an infinite chain we rewrite Eq.~(\ref{linear_eq_bulk}) using explicit expressions for stationary amplitudes:
\begin{eqnarray}
\begin{cases}
(\omega^2 - \omega_{0}^2)A + (J_1 +  J_2e^{-ik})B = 0,\\
(J_1 + J_2e^{ik})A + (\omega^2 - \omega_{0}^2)B  = 0,\\
\end{cases}
\label{linear_eq_T}
\end{eqnarray}
where $\omega_0^2 = \omega_R^2 + J_1 + J_2$, $\omega_R^2 = 2(a_2 + 6a_4v_T^2)$, $J_1 = 2b_2$, $J_2 = 2b_2+48b_4v_T^2$, $A$, $B$ are the amplitudes of the two oscillators in the unit cell [see Fig.~\ref{Suppl_scheme}(a)], and $k$ is the normalized Bloch wavenumber. The solution of this system
\begin{equation}
\omega^2 = \omega_0^2 \pm \sqrt{J_1^2 + J_2^2 + 2J_1J_2\cos(k)}
\label{tetramer_band}
\end{equation}
indicates that there is a gap in linear dispersion. Example of the dispersion diagram is plotted in Fig.~\ref{Disp_tetramer}.

\textbf{Edge states}: To analyze the existence of edge states, first, we need to ensure that the stationary displacements of the edge osicllators are equal to the displacement of the bulk oscillators $v_T$. There are two possible terminations that are shown in Figs.~\ref{Suppl_scheme}(b,c). The first equation from the general system~(\ref{dynamics_eq}) gives us the following conditions for two types of edge terminations:
\begin{equation}
\begin{sqcases}
2a_2v_T + 4a_4v_T^3 = 0,\;\;\;\text{for the edge in Fig.~\ref{Suppl_scheme}(b)},\\
2a_2v_T + 4a_4v_T^3 + 8b_2v_T + 64b_4v_T^3 = 0,\;\;\;\text{for the edge in Fig.~\ref{Suppl_scheme}(c)}.\\
\end{sqcases}
\end{equation}
Substituting the expression for $v_T$ gives us the same conditions for both types of the interfaces:
\begin{equation}
\dfrac{b_2}{a_2}=4\dfrac{b_4}{a_4}.
\label{rel_coefs}
\end{equation}

Next, we write down the equation for small oscillations on the first site, i.e. the first equation from~(\ref{linear_eq_edge}). First type of termination gives us: $(\omega^2 - \omega_{R}^2 - J_{1})A + J_{1}B = 0$; and second: $(\omega^2 - \omega_{R}^2 - J_{2})B + J_{2}Ae^{ik} = 0$. Solving these equations along with the system~(\ref{linear_eq_T}) gives us following solutions:\\
First type of termination: $k=0$, $A=B$, $\omega^2 = \omega_R^2$ or $k=\pi$, $A=-B$, $\omega^2 = \omega_R^2 + 2J_1$.\\
Second type of termination: $k=0$, $A=B$, $\omega^2 = \omega_R^2$ or $k=\pi$, $A=B$, $\omega^2 = \omega_R^2 + 2J_2$.\\
These states are not localized and their eigenfrequencies coincide with one of the boundary energies of the dispersion bands, given by~(\ref{tetramer_band}). Therefore,  the system does not possess edge states
due to the nonlinear detuning of the resonance frequencies of oscillators,.

\textbf{Interface states}: The system, however, can possess the states localized at the interface between two different tetramer phases. To analyze such situtation we perform the similar procedure as for the edge states analysis. Two types of interfaces with different coupling constants are shown in Figs.~\ref{Suppl_scheme}(d,e).\par

The  dynamics equations~Eqs.~(\ref{dynamics_eq}) written down for the interface oscillator give us the following conditions:
\begin{equation}
\begin{sqcases}
2a_2v_T + 4a_4v_T^3 + 8b_2v_T + 64b_4v_T^3 = 0,\;\;\;\text{for the interface in Fig.~\ref{Suppl_scheme}(d)}\\
2a_2v_T + 4a_4v_T^3 = 0,\;\;\;\text{for the interaface in Fig.~\ref{Suppl_scheme}(e)}.\\
\end{sqcases}
\end{equation}
Substituting the expression for $v_T$ gives us the same condition~(\ref{rel_coefs}) as for the edge oscillators for both types of the interfaces. Further analysis reveals that this condition implies that the stationary displacements of the oscillators for all considered states are equal to each other and to the displacement of the single nonlinear oscillator $v_0$, i.e.:
\begin{equation}
v_M = v_D = v_T = v_0 = \pm\sqrt{-\dfrac{a_2}{2a_4}}.
\label{stationary_states}
\end{equation}
Therefore, in a finite chain all oscillators are found in one of two possible states $\pm v_0$. This also holds for the edge and interface oscillators, which do not acquire any static shift from $v_0$. Using this condition, the expression for $J_2$ and $\omega_R^2$ can also be simplified to $J_2 = -4b_2$, and $\omega_R^2=-4a_2$.\par

Next, we write down the equations for small oscillations on the interface oscillator and on the left and right halves shown in Fig.~\ref{Suppl_scheme}(d):\\
\begin{equation}
\begin{cases}
(\omega^2 - \omega_R^2 - 2J_2)A +J_2(B_1 + B_2) = 0,\\
(\omega^2 - \omega_0^2)A_1 + B_1(J_1\exp(ik_1) + J_2) = 0,\\
(\omega^2 - \omega_0^2)B_1 + A_1(J_1\exp(-ik_1) + J_2) = 0,\\
(\omega^2 - \omega_0^2)A_2 + B_2(J_1\exp(-ik_2) + J_2) = 0,\\
(\omega^2 - \omega_0^2)B_2 + A_2(J_1\exp(ik_2) + J_2) = 0,\\
\end{cases}
\label{interface}
\end{equation}
where $A = A_1 = A_2$. Due to the mirror symmetry of the system we need to consider only antisymmetric and symmetric solutions. Moreover, since we are looking for the localized state we have $ik = ik_1 = -ik_2$ and we are interested only in solutions with $\exp(ik)>1$.\par

In the case of antisymmetric solution we have $A=0$, $B_1 = -B_2$. We immediately obtain $\omega^2 = \omega_0^2$, $\exp(ik)=-J_2/J_1 = 2$. This is the state localized at the interface between two tetramer states, with the energy that always resides exactly in the middle of the gap. For a symmetric solution we take $A\neq0$, $B_1 = B_2$. Such solutions exists with the energy $\omega^2 = \omega_0^2 + \dfrac{(J_2+J_1e^{-ik})(J_1e^{ik}-J_2)}{J_2-J_1}$, where $e^{ik} = \dfrac{3(J_2-J_1)\pm\sqrt{9(J_1-J_2)^2+4J_1J_2}}{2J_1}$, $|z|>1$. Taking into account that $J_1=2b_2$ and $J_2 = -4b_2$ we obtain $\omega^2 = \omega_0^2 - b_2(\sqrt{73}+1)$.\par

The analysis of the second type of the interface shown in Fig.~\ref{Suppl_scheme}(e) is done in the same way. The equations for this type of the interface can be obtained by replacing $J_2 \leftrightarrow J_1$ in the system~(\ref{interface}). For antisymmetric state we have $\exp(ik) = -J_1/J_2<1$, which means that this state is not localized. Therefore, for this type of the interface there is only a topologically trivial localized symmetric state.\par

Overall, by ensuring that the parameters of the system satisfy the condition~(\ref{rel_coefs}) we expect the formation of the interface state between two tetramer phases with the energy exactly in the middle of the bandgap.\par

\section{Conditions of existence and stability of the stationary states}

Taking into account the relation~(\ref{stationary_states}), we find out that existence of all considered stationary states is ensured by only one condition $\operatorname{sgn}(a_2) \ne \operatorname{sgn}(a_4)$.\par

Stability of the solutions can be checked by calculating the sign of the minimal value of $\omega^2$, which should be positive for the stable solutions. Low boundaries of the energy bands of the considered states are found from the general system~Eq.~(\ref{linear_eq_bulk}):\\
\textbf{Monomer state}:
\begin{equation}
-4(a_2 - b_2) - 4|b_2| \le \omega^2 \le -4(a_2 - b_2) + 4|b_2|.\\
\end{equation}
\textbf{Dimer state}:
\begin{equation}
-4(a_2 + 2b_2) - 8|b_2| \le \omega^2 \le -4(a_2 + 2b_2) + 8|b_2|.\\
\end{equation}
\textbf{Tetramer state}:
\begin{equation}
\begin{sqcases}
-4(a_2 + b_2/2) - 6|b_2| \le \omega^2 \le -4(a_2 + b_2/2) - 2|b_2|,\\
-4(a_2 + b_2/2) + 2|b_2| \le \omega^2 \le -4(a_2 + b_2/2) + 6|b_2|,
\end{sqcases}
\label{tetramer}
\end{equation}\\
From the stability conditions of the tetramer state~Eq.~(\ref{tetramer}) we deduce that the positive sign of $a_2$ always leads to the instability of the tetramer state. Therefore, we have $a_2<0$, $a_4>0$. Further, the ratios $b_2/a_2$ and $b_4/a_4$ have the same sign, according to the relation~Eq.~(\ref{rel_coefs}), either positive or negative. Taking this into account we can formulate the following stability criteria of the stationary states:
\begin{equation}
\begin{cases}
b_4/a_4<1/8\;\;\;\text{for the monomer state},\\
b_4/a_4>-1/16\;\;\;\text{for the dimer state},\\
-1/8<b_4/a_4<1/4\;\;\;\text{for the tetramer state}.\\
\end{cases}
\label{stability}
\end{equation}
The ranges of values of the ratio $b_4/a_4$ that correspond to the unsable solutions are illustrated in Fig.~2 in the main text with dotted grey lines for all states.

\section{Mean chiral displacement calculation}
In our calculations we consider the randomly distributed initial displacements of the oscillators. Hence, the elements of the interaction matrix that describes the dispersion of the linearized system are also  random in the beginning of  system evolution and converge to stationary values  in the limit of $t \rightarrow \infty$. Since at each moment of time the system is not periodic, the topological properties of the system cannot be derived from direct calculation of the Zak phase. Thus, we characterize  the topological properties of the system by  the so-called mean chiral displacement (MCD)~\cite{Cardano}. This method allows for determining of the topological phase of the not necessarily periodic systems. For a given system one can calculate the MCD of a freely evolved state $\mathbf{\Psi}$ as a function of time delay $\tau$ as follows:
\begin{equation}
\mathrm{MCD}(t,\tau) = \sum (\mathbf{\Gamma} \mathbf{n}) \mathbf{\Psi}(\tau+t),
\end{equation}
where $\mathbf{\Gamma}$ is the matrix of the chiral operator, $\mathbf{n}$ is the matrix of the position operator, so that\\
$\mathbf{\Gamma n}={\rm diag}(\ldots,-1,1,0,0,1,-1,2,-2,\ldots)$; and the vector of displacements $\mathbf{\Psi}=[\ldots,y_{1},y_{2},\ldots]^{T}$ is 
localized in an arbitrary unit cell for $\tau=0$. For the SSH model the value of MCD at large times $\tau$ converges to either $0$ or $0.5$ depending on the choice of the unit cell, which corresponds to the values of the Zak phase $\gamma=0$ or $\gamma=\pi$, respectively~\cite{Cardano}. However, for non-periodic systems time dependence of the MCD might not converge to any certain value. Such behaviour can be observed in Fig.~3(b) in the main text, where we plotted the limit values of MCD at $\tau=\infty$  as a function of time $t$. At small times all oscillators possess relatively large random displacements, and consequently the calculated values of the MCD are also random, i.e. the system does not exhibit nontrivial topological properties. At large times $t \gtrsim 100$ the values of MCD($\tau\to \infty,t$) converge to $0$ and $0.5$ for the left and the right sides of the chain, respectively, indicating the formation of topologically different phases.

\end{document}